\documentclass{article}
\usepackage{spconf,amsmath,graphicx}

\usepackage{enumitem}
\usepackage[letterpaper, top=1.92cm, bottom=3.2cm, left=1.92cm, right=1.92cm]{geometry}
\setlist{nosep, leftmargin=14pt}

\usepackage{mwe} 

\title{Anatomy-Guided Multitask Learning for MRI-Based Classification of Placenta Accreta Spectrum and Its Subtypes}
\name{
  Hai Jiang$^{\dagger,1}$, 
  Qiongting Liu$^{\dagger,3}$\thanks{$\dagger$ Equal Contribution}, 
  Yuanpin Zhou$^1$, 
  Jiawei Pan$^1$, 
  Ting Song$^{\ast,2}$, 
  Yao Lu$^{\ast,1}$\thanks{$\ast$ Correspondence}
}
\address{
  $^1$School of Computer Science and Engineering, Sun Yat-sen University \\
  $^2$Department of Radiology, The Third Affiliated Hospital of Guangzhou Medical University \\
  $^3$Department of Radiology, Hainan General Hospital, Hainan Medical University \\
}

\begin{document}
%
\maketitle
%
\begin{abstract}
Placenta Accreta Spectrum Disorders (PAS) pose significant risks during pregnancy, 
frequently leading to postpartum hemorrhage during cesarean deliveries and other severe clinical complications, 
with bleeding severity correlating to the degree of placental invasion. 
Consequently, accurate prenatal diagnosis of PAS and its 
subtypes—placenta accreta (PA), placenta increta (PI), and placenta percreta (PP)—is crucial. 
However, existing guidelines and methodologies predominantly focus on the presence of PAS, 
with limited research addressing subtype recognition. 
Additionally, previous multi-class diagnostic efforts have primarily relied on 
inefficient two-stage cascaded binary classification tasks. 
In this study, we propose a novel convolutional neural network (CNN) architecture designed for 
efficient one-stage multi-class diagnosis of PAS and its subtypes, based on 4,140 
magnetic resonance imaging (MRI) slices. 
Our model features two branches: 
the main classification branch utilizes a residual block architecture comprising multiple residual blocks, 
while the second branch integrates anatomical features of the uteroplacental area 
and the adjacent uterine serous layer to enhance the model's attention during classification. 
Furthermore, we implement a multitask learning strategy to leverage both branches effectively. 
Experiments conducted on a real clinical dataset demonstrate that our model achieves state-of-the-art performance.
\end{abstract}
\begin{keywords}
Placenta Accreta Spectrum, MRI, Prenatal Diagnosis, Multi-class, 
Multitask Learning
\end{keywords}
\section{Introduction}
\label{sec:intro}
  Placenta Accreta Spectrum (PAS) disorders are among the most severe complications in obstetrics and gynecology, 
  characterized by the abnormal adherence of placental villous tissue to the uterine myometrium. 
  These conditions significantly increase the risk of adverse outcomes during cesarean delivery, 
  particularly postpartum hemorrhage, which may progress to multisystem organ failure, 
  disseminated intravascular coagulation, emergency perioperative hysterectomy, 
  and even maternal death\cite{silver2018placenta,bailit2015morbidly}. 
  The severity of hemorrhage is closely associated with the depth of placental invasion, 
  with PAS classified into three subtypes: 
  placenta accreta (PA), placenta increta (PI), and placenta percreta (PP), based on the extent of myometrial involvement. 
  Given these risks, accurate prenatal diagnosis of PAS and its subtypes is critical for effective management. 
  Ultrasound (US) remains the preferred initial diagnostic modality due to its accessibility and cost-effectiveness; 
  however, its accuracy is highly dependent on the operator's expertise, resulting in variable diagnostic outcomes. 
  To address these limitations, this study utilizes T2-weighted MRI (T2WI) to evaluate the invasive characteristics of PAS. 
  The superior soft tissue resolution provided by T2WI aims to enhance diagnostic reliability in these complex cases\cite{d2014prenatal}. 


  Deep learning methods have shown significant potential in computer-aided diagnosis of PAS. 
  Lu et al.\cite{lu2023hacl}  developed an end-to-end Hierarchical Attention and Contrastive Learning Network (HACL-Net), 
  while Peng et al.\cite{peng2024prenatal} implemented a deep learning radiomics (DLR) model. 
  Wang et al.\cite{wang2024deep} incorporated the uteroplacental borderline as prior knowledge to enhance model performance. 
  Additionally, Zheng et al.\cite{zheng2024deep} utilized clinical characteristics alongside the DLR model, further improving diagnostic accuracy. 

  While recent approaches have improved PAS diagnosis performance in binary classification tasks, 
  few studies have explored subtype classification of PAS, which requires distinguishing multiple classes. 
  Most methods to date focus exclusively on binary classification, 
  with only one study\cite{zheng2024deep} addressing subtype recognition through a two-stage cascaded binary classification approach. 
  This method, however, is time-consuming and lacks efficiency, 
  as it first identifies the presence or absence of PAS and then further classifies subtypes like PI and PA. 
  Additionally, a notable gap remains in optimizing attention mechanisms within CNN models for PAS diagnosis. 
  Complex image classification tasks often contain multiple recognition patterns, 
  yet the internal feature patterns of hidden layers in CNNs are challenging to interpret and control. 

  Our contributions are as follows: 
  \begin{itemize}
    \item We propose a novel one-stage, four-class CNN model with two branches, 
          designed to classify not only non-PAS/PAS but also the three specific subtypes of PAS; 
    \item The first branch utilizes an efficient classification module with a backbone of multiple residual blocks;
    \item The second branch incorporates anatomical features of the uteroplacental area 
          and the adjacent uterine serous layer to guide the model's attention during classification; 
    \item To fully leverage both branches, each addressing distinct tasks, 
          we introduce a multitask learning strategy to enhance the CNN model's discriminative capacity.
  \end{itemize}
  Experimental results on a clinical dataset of 414 patients demonstrate that our model achieves state-of-the-art performance.
  \begin{figure*}[h!]
    \centering
    \includegraphics[width=0.98\textwidth]{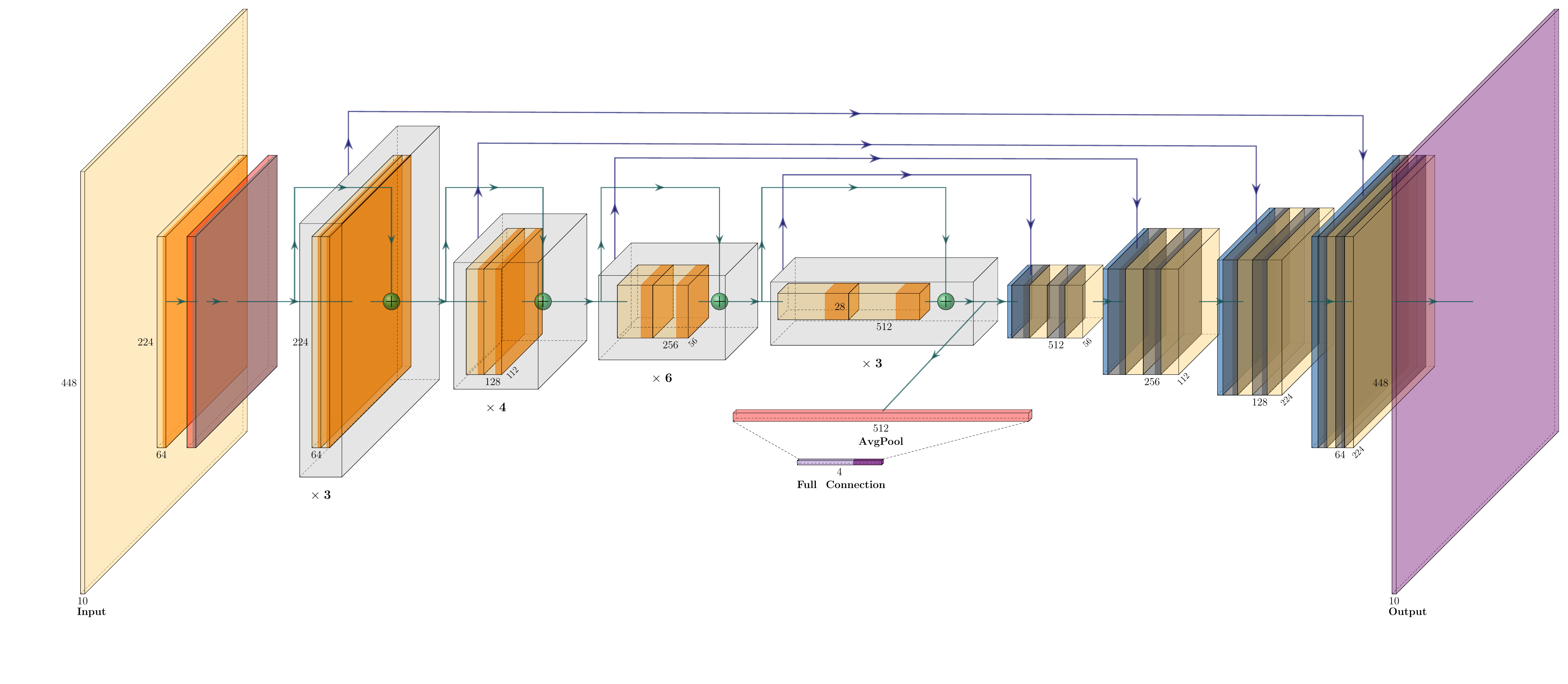}
    \caption{
      An overview of the proposed model framework. 
      The backbone of our model is the classification branch, 
      located on the left of the structure; 
      it contains four residual blocks 
      and also works as the segmentation branch's encoder. 
      The line in purple means the skip connection. 
      The ball in the green means the add operator.
    }
    \label{fig:overview}
  \end{figure*}

\section{Methodology}
\label{sec:method}
  The current study involved the 
  collection of 4,140 T2-weighted (T2WI) MRI images 
  to develop a more effective deep learning solution for classifying PAS into its three subtypes. 
  All MRI images in the dataset were sourced from 
  the Third Affiliated Hospital of Guangzhou Medical University (GZMU) over the past six years (2018 to 2023). 
  The study received approval from the Ethics Committee of the Third Affiliated Hospital of GZMU. 
  Our proposed novel method is based on a Convolutional Neural Network (CNN) and employs a multitask learning framework. 
  As illustrated in Fig. \ref{fig:overview}), 
  our model comprises two branches: the classification branch and the segmentation branch. 

  Our architecture is inspired 
  by two key tasks in deep learning 
  for medical image analysis: classification and segmentation\cite{hooper2024case}. 
  Semantic segmentation of the placenta has demonstrated high accuracy 
  in 3D ultrasound without user intervention\cite{torrents2019automatic}. 
  Yang et al.\cite{yang2018towards} introduced the first 3D fully connected networks (FCN) 
  specifically for segmenting the placenta and fetus. 
  With precise annotations from radiologists and self-adaptive parametric adjustments during training, 
  U-Net-like models can achieve remarkable results in our MRI-based placenta segmentation task.  
  In the context of deep learning (DL) in radiology, 
  image segmentation and classification are critical and complementary tasks\cite{hooper2024case}. 
  This complementary relationship is why multitask learning strategies\cite{caruana1997multitask} 
  are frequently employed in large-scale disease diagnosis\cite{cao2023large}. 

  In this section, we will first define our task mathematically. 
  Following that, we will provide detailed descriptions of each branch in our proposed model. 
  \subsection{Task Specification}
    Given an input tensor of MRI slices, the backbone (classification path) of our model 
    predicts the category among four different classes: non-PAS, PA, PI, and PP. 
    We use global average pooling (GAP) to reduce the dimensionality of the feature maps generated by the encoder, 
    followed by a softmax function to normalize the output GAP values $L=\{l_1, ..., l_C\}$, $C=4$, 
    where $C$ represents the number of classes. 
    Therefore, the loss function for class $j$ in the classification branch can be defined as: 
    \begin{equation}
      L_{cls} = -\sum_{c=1}^{C}\mathbf{1}_j\cdot\log[\frac{\exp(l_j)}{\sum_{i=1}^{C}\exp(l_i)}], 
    \end{equation}
    where $\mathbf{1}_j\in\{0, 1\}$ is an indicator of whether the class is the correct classification. 
    In the segmentation branch, for the sake of simplicity, 
    the pixel-level classification label is denoted as $p\in\{0,1\}$. 
    A value of $p=1$ indicates that the pixel is included in the image's area of interest. 
    For this purpose, we utilize the $BCEWithLogitsLoss$ function: 
    \begin{equation}
      L_{seg} = -\frac{1}{w\times h}\sum_{w,h}[p\log\hat{p} + (1-p)\log(1-\hat{p})], 
    \end{equation}
    where $\hat{p}$ indicates the prediction, and $w$ and $h$ denote the width and height of the input tensor. 
    Our proposed network, which addresses two distinct tasks, was trained simultaneously 
    to classify the MRI slices into four classes and predict the region of interest (RoI). 
    Therefore, the total loss for training the network is given by: 
    \begin{equation}
      L = L_{cls} + \lambda\cdot L_{seg}, 
    \end{equation}
    where $\lambda$ is a weighting parameter that balances the two different objective loss functions. 
    Based on preliminary experiments, we set $\lambda=1$ for our studies (more details can be found in Sec.\ref{sec:result}). 
  \subsection{Two Branches}
    The architecture of the classification branch is located on the left side of our CNN model, as illustrated in Fig.\ref{fig:overview}. 
    It takes a stack of ten consistent grayscale MRI slices as input. 
    Following a max-pooling operation, the scale of the input is halved. 
    To leverage the speed and training efficiency of ResNet\cite{he2016deep}, 
    we selected a residual block at each downsampling step to fully extract features from the ten slices. 
    With each downsampling step, the input scale is reduced by half while the number of filters (i.e., channels) is doubled. 
    Overall, there are four residual blocks in the downsampling path, which also serves as the encoder for the segmentation branch. 
    After the final residual block, which acts as the bottleneck in the middle of the network, 
    the deep features are projected into a 512-dimensional vector using the GAP operator. 
    In the last step of the classification branch, a fully connected layer is implemented to connect to four output classes. 
    
    Drawing inspiration from the empirical practices of clinical radiologists 
    and the principles of semantic segmentation tasks, 
    we incorporated anatomical features of the utereoplacental area 
    and the adjacent uterine serous layer to enhance the attention mechanism of the classification backbone. 
    The segmentation branch is positioned on the right side of the model and is a modified version of the U-Net decoder, 
    with the number of filters in each upsampling block adjusted to match the expansion rate of the encoder. 
    The number of upsampling blocks equals the number of residual blocks, 
    and the feature maps from the encoder and decoder are concatenated using skip connections. 
    Since each MRI slice includes a mask, the number of output channels is maintained to match the input, 
    allowing our model to be guided by the region of interest (RoI) in each individual slice.

\section{Experimental Results}
\label{sec:result}
  Earlier studies (Sec.\ref{sec:intro}) attempted to diagnose PAS directly from the Region of Interest (RoI), 
  and we conducted similar experiments using binary classification. 
  Following a comparable procedure, we adapted this approach for our dataset, 
  selecting ResNet18 as the feature extractor. 
  Each MRI slice was combined with its corresponding mask, 
  treating each grayscale image as a single channel and using the mask as prior knowledge.
  All classification metrics, including AUC, accuracy, 
  and specificity, consistently approached a perfect score of $\mathbf{1.00}$. 

  The proposed architecture was used for the one-stage 
  task of classifying the PAS and the subtype of spectrum. 
  \subsection{Dataset and Preprocessing}
    The experiments are conducted on a clinical dataset collected from a hospital 
    between January 2018 and June 2023. 
    This dataset includes 414 patients, each with ten MRI slices focused on the placental region, totaling 4,140 slices. 
    Among these, 238 patients were confirmed to have PAS through surgery and pathology (PA, $n=85$; PI, $n=88$; PP, $n=65$). 
    All patients underwent preoperative pelvic MRI scans between 28 and 34 weeks of gestation. 
    A skilled radiologist segmented the placental area and the adjacent uterine serous layer in the sagittal T2WI images. 
    The placenta and the adjacent uterine serous layer area of the placenta 
    were segmented in the sagittal T2WI by an adept radiologist. 
    For comparison, the remaining 176 cases, which were benign and unannotated, 
    were automatically segmented using nn-UNet\cite{isensee2021nnu}, 
    with the resulting masks precisely refined by the same radiologist. 
    The input volume was standardized to $448\times448\times10$, 
    except for models such as MaxViT, SWIN Transformer, and Vision Transformer, 
    which were normalized to $224\times224\times10$. 
  \subsection{Implementaion Details}
    The proposed CNN model is trained on an NVIDIA GeForce GTX 1080 Ti GPU with 12GB of RAM. 
    A stratified 5-fold cross-validation approach, implemented using the $scikit-learn$ 
    library to maintain consistent class distribution across folds, is employed. 
    Each fold is trained for 500 epochs, and the metrics for each fold, along with their average values, were reported. 
    The total training time was approximately 11 hours. 
    The model is optimized using the $Adam$ optimizer with a  
    $learning\ rate$ of $0.0001$, $batch\ size$ of $16$, and $\lambda$ set to $1.0$. 
    The learning rate is reduced by a factor of $\gamma = 0.5$ after each step. 
    Additional hyperparameters included $n_{in}=10$ (the number of slices in the model input) 
    and $n_f=16$ (the basic filter number of the encoder). 
  \subsection{Results and Comparison}
    \begin{table}[h]
      \label{tab1}
      \begin{center}
        \caption{
        Performance comparison between our proposed method and existing methods for image classification. 
        AUC represents the area under the ROC curve. 
        Each metric value is averaged over 5 folds. Bolded results indicate the best performance. 
        }
        \begin{tabular}{c|c|c|c}
          \hline
          \multicolumn{2}{l|}{\textbf{Method}} & \textbf{AUC} & \textbf{Accuracy} \\
          \hline
          \multicolumn{2}{l|}{AlexNet}             & 0.76868 & 0.44064 \\
          \multicolumn{2}{l|}{ConvNext}            & 0.63064 & 0.34220 \\
          \multicolumn{2}{l|}{DenseNet}            & 0.74680 & 0.44986 \\
          \multicolumn{2}{l|}{EfficientNet}        & 0.76414 & 0.45000 \\
          \multicolumn{2}{l|}{GoogleNet}           & 0.77578 & 0.48244 \\
          \multicolumn{2}{l|}{Inception}           & 0.78320 & 0.46124 \\
          \multicolumn{2}{l|}{MaxViT}              & 0.77254 & \textbf{0.48790} \\
          \multicolumn{2}{l|}{MNASNet}             & 0.58172 & 0.25000 \\
          \multicolumn{2}{l|}{MobileNet}           & 0.76010 & 0.43988 \\
          \multicolumn{2}{l|}{ResNet}              & 0.77276 & 0.39982 \\
          \multicolumn{2}{l|}{ResNext}             & 0.76330 & 0.43056 \\
          \multicolumn{2}{l|}{ShuffleNet}          & 0.72376 & 0.42228 \\
          \multicolumn{2}{l|}{SqueezeNet}          & 0.62728 & 0.33440 \\
          \multicolumn{2}{l|}{SWIN-Transformer}    & 0.65204 & 0.31842 \\
          \multicolumn{2}{l|}{VGG}                 & 0.75666 & 0.46266 \\
          \multicolumn{2}{l|}{Vision-Transformer}  & 0.64890 & 0.34812 \\
          \multicolumn{2}{l|}{Wide-ResNet}         & 0.76360 & 0.39366 \\
          \multicolumn{2}{l|}{Proposed}            & \textbf{0.80150} & 0.48231 \\
          \hline
        \end{tabular}
      \end{center}
    \end{table}
    \begin{table}[h]
      \label{tab2}
      \begin{center}
        \caption{Ablation study with different branch.}
        \begin{tabular}{c|c|c}
          \hline
          \textbf{Backbone} & \textbf{Segmentation Branch} & \textbf{AUC} \\
          \hline
          $\surd$ & & 0.79114 \\
          $\surd$ & $\surd$ & \textbf{0.80150} \\
          \hline
        \end{tabular}
      \end{center}
    \end{table}
    \begin{table}[h]
      \label{tab3}
      \begin{center}
        \caption{Ablation study with different balance parameter.}
        \begin{tabular}{c|c|c}
          \hline
          \textbf{Balance parameter $\lambda$} & \textbf{AUC} & \textbf{Accuracy} \\ 
          \hline
          0.5 & 0.77964 &  0.45774 \\
          0.9 & 0.78702 &  0.46224 \\
          1.0 & \textbf{0.80150} &  \textbf{0.48231} \\
          \hline
        \end{tabular}
      \end{center}
    \end{table}
    Table 1 presents the performance of our proposed network compared to other common architectures. 
    To evaluate classification ability, we used AUC (area under the ROC curve) and accuracy as metrics, 
    both obtained through the \emph{TorchMetrics} package. 
    Given the equal importance of each class, we selected Macro-AUC (one-vs-rest) as our evaluation metric. 
    As shown in Table1, our model achieves the highest AUC of 80.150\%. 
    Table 2 also illustrates the contribution of each branch in our model. 
    Notably, using only the backbone without the segmentation branch results in an AUC of less than 80.000\%. 
    Additionally, when comparing the AUC of the backbone to other methods in Table 1, 
    it is evident that our classification branch is highly efficient. 
    The table also includes an experiment assessing the balance parameter $\lambda$ used in the multitask procedure, 
    demonstrating that the optimal results are achieved when $\lambda=1$. 
\section{Conclusion}
\label{sec:conclusion}
  This work presents a novel and efficient one-stage CNN model for four-class classification, 
  featuring two branches that differentiate between non-PAS, PAS, and its three specific subtypes. 
  The first branch employs a classification module built on a backbone of multiple residual blocks. 
  The second branch integrates anatomical features of the uteroplacental area and 
  the adjacent uterine serous layer to direct the model's attention during classification. 
  To fully leverage the capabilities of both branches, each addressing distinct tasks, 
  we introduce a multitask learning strategy aimed at enhancing the CNN model's discriminative capacity. 
  Experimental results on a clinical dataset of 416 patients demonstrate that our model achieves state-of-the-art performance, 
  showcasing significant potential for clinical applications.
\section{Acknowledgments}
  This work was supported by National Key Research and Development Program of China [No. 2023YFE0204300], 
  National Natural Science Foundation of China 
  [No. 82441027, 62371476, 82371917, 82102130], 
  Guangzhou Science and Technology Bureau [No. 2023B03J1237], 
  R\&D Program of Pazhou Lab (HuangPu) [No. 2023K0606], 
  Health Research Major Projects of Hunan Health Commission [No. W20241010], 
  Guangdong Province Key Laboratory of Computational Science at the Sun Yat-sen University under Grant 2020B1212060032.
\bibliographystyle{IEEEbib}
\bibliography{refs}
\end{document}